\documentclass[11pt]{article}
 \pdfoutput=1
\usepackage{url}
\usepackage{hyperref}
\usepackage{amssymb}
\usepackage{xcolor}
\usepackage[pdftex]{graphicx}
\usepackage{epsfig}
\usepackage{cite}

\newcommand{\AmS}{{\protect\the\textfont
  A\kern-.1667em\lower.5ex\hbox{M}\kern-.125emS}}
\newcommand{\ba}{\begin{array}}
\newcommand{\ea}{\end{array}}

\def\hg{{\rm h}}
\def\rcg{{\rm rec}}

\def\clg{{\rm c}}
\def\qcdg{{\rm QCD}}
\def\beq{\begin{equation}}
\def\eeq{\end{equation}}

\def\beq{\begin{equation}}   
\def\eeq{\end{equation}}

\def\bea{\begin{eqnarray}}
\def\eea{\end{eqnarray}}

\begin{document}
\begin{titlepage}

\begin{flushright}
 IFIC/20-28, FTUV-20-06-11\\
\end{flushright}

\begin{center}
\vspace{2.7cm}
{\Large{\bf
Cosmological analogies in the search for new physics\newline
\vskip 0.05cm
in high-energy collisions}}
\end{center}

\vspace{1cm}

\begin{center}

{\bf Miguel-Angel Sanchis-Lozano$^{\rm a,\ast}$,  
Edward K. Sarkisyan-Grinbaum$^{\rm b,c,\ast\ast}$, 
Juan-Luis Domenech-Garret$^{\rm d,\dagger}$ and
Nicolas Sanchis-Gual $^{\rm e,\ddagger}$
\vspace{1.5cm}\\
\it 

\it $^{\rm a}$ Instituto de F\'{\i}sica
Corpuscular (IFIC) and Departamento de F\'{\i}sica Te\'orica \\
\it Centro Mixto Universitat de Val\`encia-CSIC, Dr. Moliner 50, E-46100 Burjassot, Spain}
\\ 
\it $^{\rm b}$ Experimental Physics Department, CERN, 1211 Geneva 23, 
Switzerland\\
\it $^{\rm c}$ Department of Physics, The University of Texas at Arlington, 
TX 
76019, USA
\\
\it $^{\rm d}$ Departamento de F\'{\i}sica A.I.A.N., 
Universidad Polit\'ecnica de Madrid, E-28040 Madrid, Spain
\\
\it $^{\rm e}$ Centro de Astrof\'{i}sica e Gravita\c{c}\~{a}o - CENTRA,
Departamento de F\'{i}sica, Instituto Superior T\'{e}cnico - IST, Universidade de Lisboa
- UL, Avenida Rovisco Pais 1, 1049-001 Lisboa, Portugal

\end{center}

\vspace{0.5cm}

\begin{abstract} 

In this paper, analogies between multiparticle production in high-energy
collisions and the time evolution of the early universe are discussed.
A common explanation is put forward under the assumption of an
unconventional
early state: a rapidly expanding universe before recombination (last
scattering
surface), followed by the CMB, later evolving up to present days, versus 
the
formation of hidden/dark states in hadronic collisions followed by a
conventional QCD parton shower yielding final-state particles.
In particular, long-range angular correlations are considered pointing out
deep
connections between the two physical cases potentially useful for the
discovery
of new physics.
 \end{abstract}

\begin{center}

\today

\end{center}
\vskip 2.2cm

\begin{small}
\noindent 
$^\ast$E-mail address: Miguel.Angel.Sanchis@ific.uv.es \\   
$^{\ast\ast}$E-mail address: Edward.Sarkisyan-Grinbaum@cern.ch \\
$^\dagger$E-mail address: domenech.garret@upm.es \\
$^\ddagger$E-mail address: nicolas.sanchis@tecnico.ulisboa.pt
\end{small}

\end{titlepage}


\section{Introduction}

The study of correlations
 has decisively contributed to 
 the advancement of 
 scientific knowledge in all branches of physics, 
 from condensed matter and quantum 
 information to particle physics and 
cosmology. 
In the latter case, the systematic study of correlations was considered 
  in the context of the large structure of the universe 
  \cite{Peebles1980}. In this sense, the homogeneity, 
thus long-range angular correlations 
of the cosmic microwave background (CMB) across the 
sky seen by the WMAP and Planck missions 
\cite{Spergel:2003,Ade:2013kta}, strongly supports 
   an inflationary era of the early universe 
 \cite{Guth,Linde}. This proposal gave birth to
a new paradigm in astrophysics and cosmology ultimately 
leading to the Standard Cosmological Model ($\Lambda$CDM). 

Similarly, the study of angular correlations in  
 high-energy
collisions has 
traditionally been a 
common tool to understand
multiparticle production in  
 particle
collisions
 beginning with early studies of cosmic rays through to current
investigations at  
the LHC. 
  In earlier 
 papers we show
 that one consequence of the production of
a new still unknown stage of matter in high-energy hadronic collisions 
is to enhance
 long-range angular correlations among final-state particles  
\cite{Sanchis-Lozano:2009,Sanchis-Lozano:2018wpz,Sanchis-Lozano:2018mur}. 
 This conclusion
bears a  certain resemblance with the observed small 
temperature fluctuations 
of the CMB requiring
an inflationary period  right after the Big Bang. 

As is well known, analogies between different fields of knowledge have 
traditionally played an important role in the advance of science. A 
paradigmatic example in physics is provided by the analogy between 
superconductivity in condensed matter physics, and the vacuum screening 
currents leading to the Higgs mechanism in elementary particle physics 
\cite{Aitchinson}. Although the physical origin may be totally distinct 
(Cooper electron pairs versus a Higgs quantum field current), a mapping 
can be established between the equations governing both processes, as well 
as some specific relations between the theory parameters. Actually, such 
an analogy proved to be a useful guide for getting a deeper understanding 
of the origin of mass and further developments of the electroweak theory.

The main goal of this paper is showing  an analogy between 
the cosmic evolution of the universe and multiparticle production in 
high-energy collisions, as well as its consequences as a new way of 
hunting hidden/dark matter at the LHC and other future facilities. A 
caveat is in order however: in the former case there is only one universe 
(ours) to be observed, while in the latter a large number of independent 
collisions are statistically considered altogether. This difference should 
not alter the main consequences of our analogy.

\section{Angular correlations in modern cosmology}

According to the firstly postulated Big Bang Theory, the angular scale of
the horizon on the last
scattering surface  (when  the  CMB  was  emitted)  should  be
$\theta \simeq 1^\circ$ \cite{Padmanabhan}.   This
implies  that  strong  temperature inhomogeneities should show up above
this scale in contrast to
real measurements which reveal an extremely isotropic and homogeneous
microwave background. To solve this problem,
an inflationary era in the very early universe was proposed, flattening
all fluctuations up to
very large opening angles covering the entire celestial sphere.


On the other hand, the emergence of large-scale features in the CMB are attributed 
to density fluctuations in the early universe evolving into the 
large-scale structure as we see today. In fact, 
it is common wisdom that such small 
temperature fluctuations 
 (of the order of $10^{-5}$~K) are the seeds of the current 
observed 
galaxy distributions, galaxy clusters
and higher macro-structures of our universe. 

Two categories of temperature fluctuations observed in the CMB
can be distinguished according to the universal time evolution:
(a) {\em primary} anisotropies, prior to decoupling, and 
(b) {\em secondary} anisotropies
developing as the CMB propagates from the surface of the last scattering 
to the 
observer. The former include temperature inhomogeneities due to photon 
propagation under
  metric fluctuations, the so-called Sachs-Wolfe (SW) effect. This
effect
shows up at rather large angles, i.e. for $\theta \gg  1^\circ$,  
where $\theta$ stands for the angular separation 
of different directions in the present sky. Moreover, the primitive plasma
also underwent acoustic oscillations prior to decoupling associated
to a typical angular scale $\theta \lesssim  1^\circ$.
  
On the other hand, once photons decoupled
from baryons after recombination, the CMB propagated through a large 
structure where the gravitational and 
inter-cluster gas which are not be necessarily isotropic nor 
homogeneous on small
spatial scales. Examples of such secondary
anisotropies of the CMB include the 
  Sunyaev-Zeldovich 
 effect due to 
thermal electrons, 
and the integrated 
   SW effect, induced by the time variation of 
gravitational potentials.
 These effects are mainly
expected to produce temperature fluctuations on 
arc-minute scales. In this work, we shall consider them altogether
under a common parametrization of very-short-range correlations.

Let us emphasize that what matters in our analogy on angular correlations
is the existence of two well differentiated
steps in the evolution of the universe, before and after recombination. 
Therefore, rather than modeling an inflationary epoch
in the primitive universe, 
we sill assume a linearly expanding universe whose scale factor reads:  
$a(t)=t/t_f$, where $t$ stands for the universal time and $t_f$ for the
time elapsed since the Big Bang to present. 
In Refs. \cite{Melia:2012xj,Melia:2017fcx} a model of this kind
was proposed and developed to explain the observed correlations of the CMB.
Then, the maximum fluctuation size at any given time $t$ can be estimated as
$\lambda_{\rm max}=2\pi R(t)$ with $R$ as the cosmic horizon radius.
   
Following the reasoning of \cite{Melia:2012xj}, 
the comoving distance to the last scattering surface 
(at recombination time $t_{\rcg}$) reads
\beq\label{eq:thetamax1}
r_{\rcg}\ =\ ct_{f}\int_{t_{\rcg}}^{t_{f}}
\frac{dt}{t}=ct_{f}\ 
\ln{\biggl[\frac{t_{f}}{t_{\rcg}}\biggr]}\, .
\eeq 
Thus the maximum angular 
size $\theta_0$ of fluctuations associated to
the CMB emitted at $t_{\rcg}$ is given by
\beq
\theta_0=\frac{\lambda_{\rm max}}{R(t_{\rcg})}\, ,
\eeq
where
\beq\label{eq:thetamax2}
R(t_{\rcg})\ =\ a(t_{\rcg})\ r_{\rcg}\ =\ 
ct_{\rcg}\ln{\biggl[\frac{t_f}{t_{\rcg}}\biggr]}\, .
\eeq 
Finally, one gets
\beq\label{eq:thetamaxfinal}
\theta_0\ \sim\ \frac{2\pi}{\ln{[t_f/t_{\rcg}]}}\ \simeq\ \frac{\pi}{5}\, ,
\eeq
where the numerical estimate corresponds to $t_f=13.8$~Gyr and 
$t_{\rcg}=3.8 \times 10^5$~yr. This value roughly
agrees with the curve determined 
from Planck data ($\simeq\ \pi/3$) as shown in \cite{Melia:2012xj}. 
 Let us remark  that Eq. (\ref{eq:thetamaxfinal})
is considered here as a simple indicator of long-range angular
correlations in the CMB, to be later ``translated'' to high-energy 
hadronic collisions.

\section{New physics from azimuthal correlations in high-energy 
collisions}

Long-range angular correlations (both in pseudorapidity and azimuth) 
also show up in multiparticle production in 
 both
$pp$ and heavy-ion collisions \cite{CollEff-rev}. From 
general arguments based
on causality, such long-range correlations can be traced back to the
very early times after the primary parton-parton hadronic interactions.
As stressed in Ref.\cite{Sanchis-Lozano:2009}, if the parton shower 
were to be altered by the presence 
of a non-conventional state of matter,  
  final-state particle correlations should be sensitive to it.

We focus on strongly interacting dark sectors arising in a wide 
variety of new physics scenarios
 like 
 e.g. 
 the Hidden Valley \cite{Strassler:2006im,Kang:2008ea,Strassler:2008fv}. 
 Hidden Valley models predict 
the 
existence of a hidden/dark sector 
connected to the Standard Model (SM) of
particle physics through heavy mediators via different 
mechanisms (tree-level, higher loop diagrams).  One of the most 
interesting situations from a phenomenological viewpoint 
corresponds to a QCD-inspired  
scenario with a hidden (running) coupling constant and a
confinement scale $\Lambda_\hg$. Then the hidden/dark quarks, which could 
be 
much lighter than the  energy scale 
set by the heavy mediator $M_\hg$, can 
form 
bound states at $\Lambda_\hg$ as hadrons in 
QCD.  Usually the masses of the hidden sector particles are assumed to lie 
below the electroweak scale while the mediators may have TeV-scale masses.
Therefore, it seems quite natural to expect large hierarchies between
 $\Lambda_\hg$ and the hidden quark masses 
$m_\hg$ \cite{Cohen:2017pzm}: 
\beq\label{eq:hier}
 m_\hg\  \sim\ \Lambda_\hg\ \ll\ M_\hg\, .
\eeq
where the condition on the relative value of
$\Lambda_\hg \ \sim\ m_\hg$ can be taken quite loosely without changing
the main conclusions of this paper.

In our numerical estimates 
  we assume that the strongly coupled hidden sector includes 
some families of hidden quarks that bind into 
hidden hadrons at energies below
  $\Lambda_{\hg} \sim O(10)$ GeV, 
 playing a similar role as $\Lambda_{\qcdg}$
in the conventional strong interaction.
  Such a simplified picture is compatible with the expected walking 
behaviour requiring a strong coupling 
over a large energy window before reaching $\Lambda_\hg$, 
thereby yielding a large number of hidden partons and 
 ultimately high multiplicity events where the primary energy is 
democratically shared by final-state particles.  

On the other hand, the SM sector could feebly couple
to the hidden sector 
   (and the equivalent hadronic 
   hidden particles and states) with a
substantial freedom in the form of the portal interaction:
 via a tree-level neutral $Z'$ or higher-order loops involving particles 
with charges of both SM and hidden sectors. 
In fact, for some (reasonable) values of the parameter space in  hidden 
valley 
models, hidden particles
can promptly decay back into SM particles, 
altering the subsequent conventional parton shower \cite{Strassler:2008fv}
and yielding (among others \cite{Cohen:2017pzm}) observable consequences, 
e.g. extremely long-range
correlations especially in azimuthal space \cite{Sanchis-Lozano:2018wpz}. 

The maximum length, i.e. time \footnote{Hereafter natural units, 
$c=\hbar=1$, 
will be used unless otherwise stated.}, of a parton shower initiated by 
a parton, down to a low virtuality scale $Q_0$, can be estimated as 
\cite{Renk:2010mf}
\beq\label{eq:cascade2}
L_{\rm max}\ \simeq\ \frac{E}{Q_0^2}\, ,
\eeq
where $E$ stands for the typical energy of the parton cascade; $Q_0$ is 
expected to be
of the 
order 
of 
 $\Lambda_{\qcdg}$ for a conventional QCD cascade and of the 
order
of 
$\Lambda_\hg$ for a dark cascade.

\begin{figure}[t!] 
\begin{center}
\includegraphics[scale=0.42]{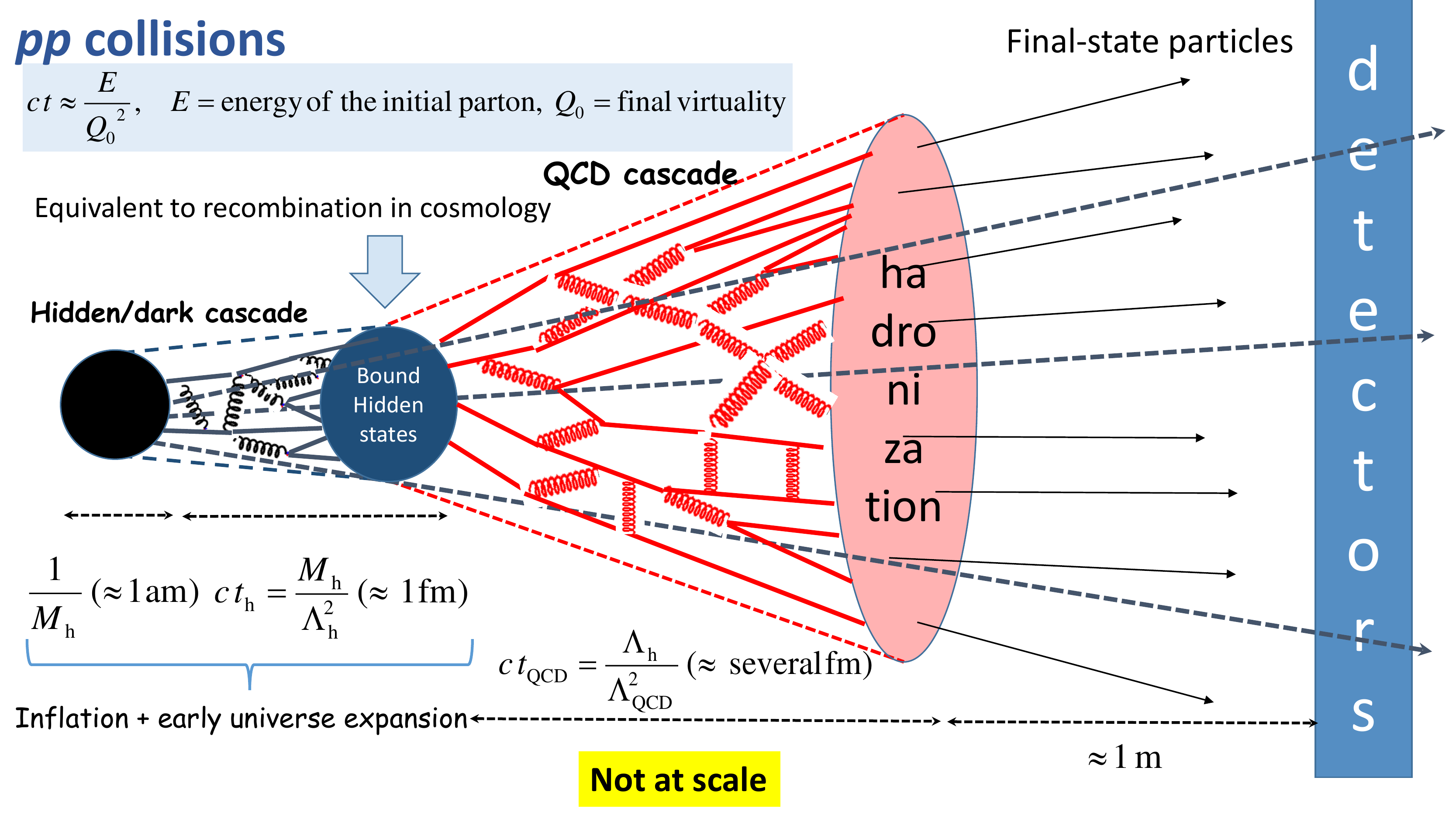}
\caption{Time evolution of the parton shower in high-energy collisions 
with formation of an initial hidden stage of matter evolving into bound states (equivalent to recombination
in cosmology) after a time $t$, followed by a decay back to SM quarks and gluon partons
lasting about $t_{\qcdg}$. There could also be hidden particles 
not decaying back to SM and therefore not detected, represented 
as dashed arrows.}
\label{fig:1}
\end{center}
\vspace*{-0.3cm}
\end{figure}

In Fig.~\ref{fig:1}, we show pictorially the foreseen evolution after a 
primary 
hard parton-parton interaction
producing a hidden shower as a first stage of the cascade ultimately 
yielding final-state particles
via a QCD parton cascade. Three steps can be distinguished:
\begin{itemize}
\item Production of heavy mediators of mass ${\cal O}(10^3)$ GeV in the 
primary partonic collision. This
is assumed to occur at a tiny
fraction of a second ($\simeq 1$~am/$c$), fixed by the 
energy scale $M_\hg$.
 \item Hidden shower and formation of hidden bound states (equivalent to
recombination in cosmology), at typical time $t_\hg=M_\hg/Q^2$, 
where $Q^2$ stands for the virtuality 
of the hidden shower. 
Assuming that $Q^2$ is of the 
  order of
 $\Lambda_\hg^2 \simeq 100$ GeV$^2$, one 
gets a time scale of the 
 order of 1 fm/$c$.
\item Once the hidden bound states 
 (or a part of them)
decay back to QCD partons (quarks and 
gluons), a ``conventional''
cascade takes place with typical time 
$t_{\qcdg}=\Lambda_\hg/\Lambda_{\qcdg}^2 \sim $ several fm$/c$, 
where we have assumed that
the typical energy of the now conventional parton shower is provided by 
$\Lambda_\hg \sim m_\hg$.
\end{itemize}


\section{Cosmological analogies}

Since the successful running of the heavy-ion program at the LHC,  
it has become popular to compare the evolution 
of the universe, some seconds after the Big Bang, with the formation 
of very dense matter at high temperature (presumably
forming a soup of quarks and gluons) in hadronic collisions. 
It has even become customary 
talking somewhat loosely about a ``little Big Bang'' at the LHC. 
Moreover, such a parallelism between the space-time developments of heavy-ion 
collisions and the early universe
has been considered beyond purely outreach purposes as a source
of physical inspiration, see e.g. Ref. \cite{Sorensen:2008dm}. In a recent
paper \cite{Li:2020vav}, the authors established a correspondence between
high-energy collions at future $e^+e^-$ colliders (ILC and CLIC)
and the CMB map.  

It should be mentioned that physics underlying angular correlations is 
completely different in the two cases: the cosmological evolution is 
fundamentally described by General Relativity whereas the parton cascade 
evolution in high-energy collisions is essentially governed by 
conventional or hidden strong interaction dynamics. However, on the one 
hand they share a common fact put forward to explain long-range 
correlations: a rapidly growing initial-state. On the other hand, the 
typical values of the angular scales are (by coincidence) quite similar as 
we shall see. Indeed, primary long-range angular correlations are of the 
order of one radian, while secondary scales lie one order of 
magnitude or more below. Such a numerical concordance of scales, together 
with the fact that the time evolution in both cases is not continuous but 
rather involves different well-defined steps, makes a connection between 
the two cases. 

 
Now, turning to 
 high-energy collisions
  and
  naively applying the same expression 
  (\ref{eq:thetamax2}) used for 
  a particular cosmological model, 
  setting $t_\hg=M_\hg/\Lambda_\hg^2$
, $t_{\qcdg}=\Lambda_\hg/\Lambda_{\qcdg}^2$ and
$t_f \simeq t_\hg+t_{\qcdg}$ 
(see Fig. \ref{fig:1}), we get for the maximal azimuthal 
correlation 
angle
\beq\label{eq:phimax}
\phi_{0}\ \simeq\ \frac{2\pi}{\ln{[t_f/t_\hg]}}\ \simeq\ \pi\,
\eeq
for $\Lambda_\hg \simeq 10$ GeV and $M_\hg \simeq 1$ TeV as reference values.  
As already commented, this result points at very long-range correlations 
emerging in a ``universe'' under the above-mentioned evolution 
conditions \footnote{Of course, the actual situation in high-energy 
particle 
 collisions at colliders  
is not the same as 
in an expanding universe, where space itself is being created as the 
expansion goes on. Nevertheless, 
one can still keep in mind the 
picture of a growing particle horizon to be 
identified somehow with the radius of a growing
sphere containing the developing parton cascade inside.}.  
Such an order-of-magnitude estimate is in agreement with our earlier 
estimates \cite{Sanchis-Lozano:2018mur} about the expected long-range 
correlation length in azimuthal space arising from new physics. Here we 
explore further the analogies between cosmic evolution and hadronic 
evolution under the presence of a hidden sector on top of the QCD shower.
Note that temperature fluctuations are supposed to be 
the seeds of the current observed 
galaxy distributions, galaxy clusters
and higher macro-structures of our universe. 
Similarly, hidden (bound) states would act as the seeds 
of clusters (jets in a broad sense) of final-state particles.

The 2-point correlation function is defined in many different (though 
essentially related) ways in the literature. For example, in the case of 
the temperature correlations of the CMB seen in directions $\vec{n}_1$ and
 $\vec{n}_2$ of the sky, the 2-point correlation function can be written as the 
ensemble-average product:
\beq
C_2(\cos{\Delta\theta}) = \langle T(\vec{n}_1)\ T(\vec{n}_2) \rangle \, ,
\eeq
where isotropy and homogeneity of space have been assumed, and 
$\cos{\Delta\theta}= \vec{n}_1 \cdot \vec{n}_2$. 
It  measures the conditional probability
of having two CMB temperatures in 
the sky plane differing by $\Delta\theta=\theta_1-\theta_2$.  Furthermore, the 
3-point angular correlation
function is defined as \cite{Gaztanaga:2003hs} 
\beq
C_3(\cos{\Delta\theta_{12}},\cos{\Delta\theta_{13}}) = \langle 
T(\vec{n}_1)\ T(\vec{n}_2)\ T(\vec{n}_3) \rangle \, ,
\eeq
where now three different directions in the sky are labelled by three vectors 
$\vec{n_i}$, $i=1,2,3$,   
and $\Delta \theta_{12}=\theta_1-\theta_2$, 
$\Delta\theta_{13}=\theta_1-\theta_3$.
Note that actually only two angular differences are independent, here 
chosen $\Delta\theta_{12}$ and $\Delta\theta_{13}$, so that  
$\Delta\theta_{23} = 
\theta_2-\theta_3 = \Delta\theta_{13}-\Delta\theta_{12}$.

Notice
 that already since some time ago 
 the study of 3-point correlations has 
  been 
 recognized as a powerful 
probe of the origin and evolution of structures of 
 the 
universe,  see e.g. \cite{Gangui:1993tt,Chen:2005ev,Komatsu:2009kd}. 
 Specifically, non-Gaussian contributions to cosmological correlations 
 should play a leading role in understanding 
 the physics of the early universe, when primordial seeds for 
 large-scale structures were created, 
and their subsequent growth at later times. 
Interestingly, the correlation function method was recently
proposed in Ref. \cite{Green:2020whw} to distinguish between quantum and
classical primordial fluctuations in a sense close to our
consideration.

In  
  high-energy 
   collisions, the
 2-particle correlation function 
$C_2(\phi_1,\phi_2)$
is similarly defined where $\phi_i$ stands for the azimuthal emission 
angle of particle $i$ measured 
on the transverse plane of a reference frame whose $z$ 
axis corresponds to the beams direction. Under rotation symmetry, the 
2-point correlation function 
actually depends only on the azimuthal difference $\Delta 
\phi=\phi_1-\phi_2$, 
i.e. $C_2(\phi_1,\phi_2)=C_2(\Delta \phi)$.

Again, higher-order correlations are 
useful as well to get a deeper insight
into multiparticle dynamics in hadronic collisions \cite{book}. 
A dependence of the correlations on angular differences is
expected too, 
 e.g. the 3-particle azimuthal correlation function $C_3(\Delta 
\phi_{12},\phi_{13})$, where 
$\Delta\phi_{12}=\phi_1-\phi_2$, $\Delta \phi_{13}=\phi_1-\phi_3$. 
Let us point out that 3-point/particle correlations constitute the 
lowest-order statistical tool to check 
the non-Gaussianity 
of distributions. Furthermore, they can place strong constraints on 
underlying clustering structures, thereby
becoming specially suited to uncover new physics in multiparticle 
production in 
 particle 
collisions as
stressed in \cite{Sanchis-Lozano:2018}. Generally speaking, 
3-particle angular correlations may suggest 
    the formation of primary clusters \cite{Wang:2009kd}; larger 
  cluster sizes imply stronger 3-particle correlations.

\begin{figure}[t!]
\begin{center}
\includegraphics[scale=0.69]{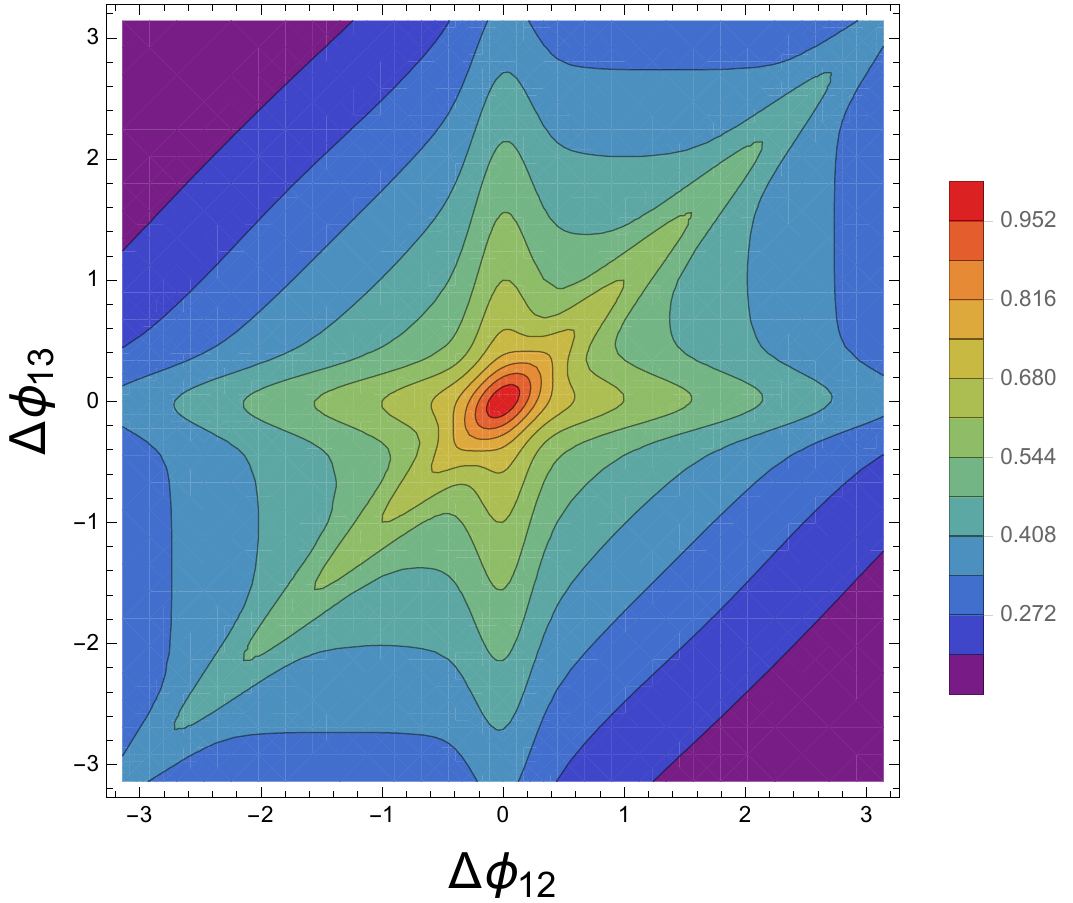}
\includegraphics[scale=0.69]{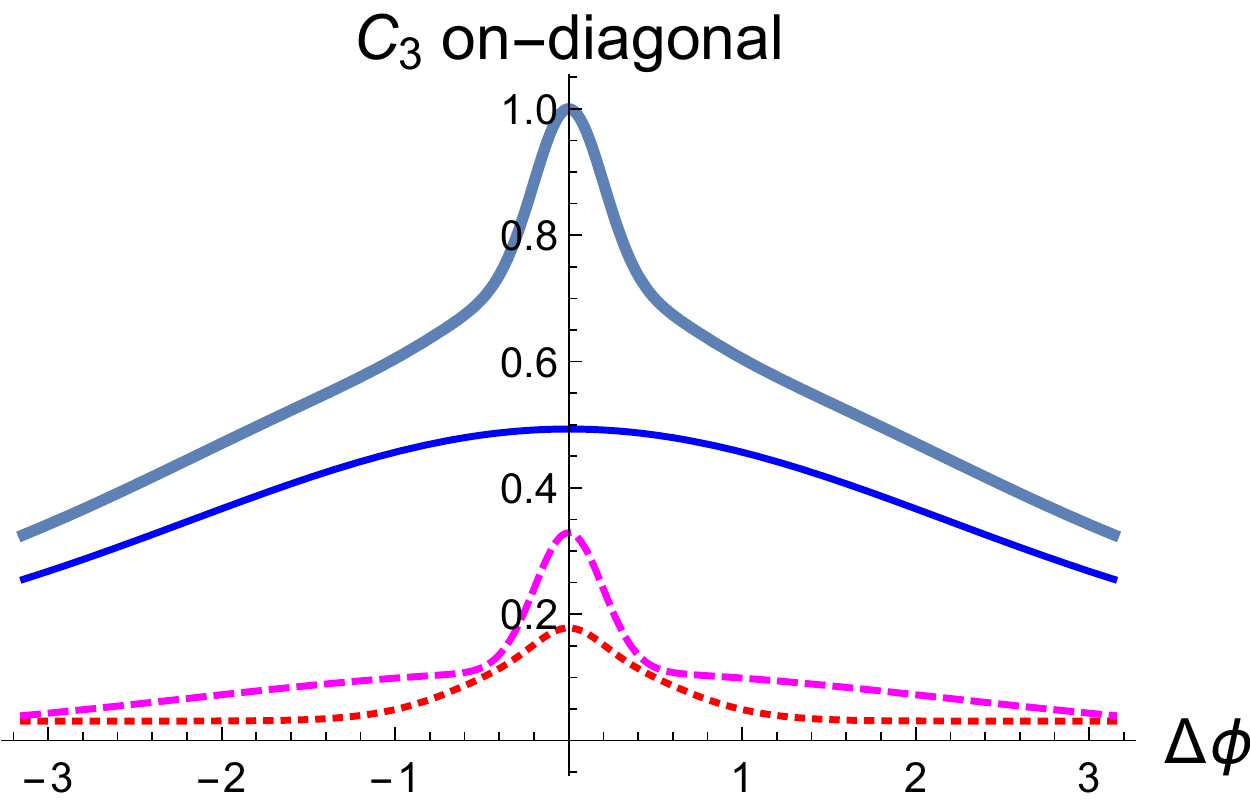}
\caption{ {\em Left}: 
 Contour plot of the 3-particle correlation function $C_3(\Delta 
\phi_{12},\Delta \phi_{13})$ of a
3-step cascade with an initial long-range contribution from a hidden sector; 
{\em Right}: Diagonal projection 
of $C _3(\Delta \phi_{12},\Delta \phi_{13})$, where the peak at 
$\Delta \phi_1= \Delta \phi_2=0$ is normalized to unity. 
The dotted (red), dashed (magenta) and thin solid (blue) curves show
the contributions from one, two and three
hidden particles, respectively. The weighted sum is shown by the
 thick (turquoise) curve.
Plots are taken from our work \cite{Sanchis-Lozano:2018wpz}.}
\label{fig:contour-proj-phi}
\end{center}
\vspace*{-0.3cm}
\end{figure}

In our approach to multiparticle production, correlations are modeled by 
using Gaussian distributions for either cluster and final state particle 
production in 
 high-energy
collisions \cite{Sanchis-Lozano:2016qdaw}.
 Thereby we make use of Gaussian widths 
to parametrize the typical correlation lengths in the different steps of 
hadron production.

As shown in \cite{Sanchis-Lozano:2018wpz}, the 3-particle 
correlation function can be 
written as: \beq\label{eq:allfunctions}
C_3(\Delta \phi_{12},\Delta \phi_{13})= \frac{1}{\langle N_{\hg}\rangle^2}
h^{(1)}(\Delta 
\phi_{12},\Delta \phi_{13})+
\frac{1}{\langle N_\hg\rangle}h^{(2)}(\Delta \phi_{12},\Delta 
\phi_{13})+h^{(3)}(\Delta 
\phi_{12},\Delta \phi_{13}),
\eeq
where each term on the r.h.s. represents the correlations due to one, two 
and three initial 
sources of hidden particles, indicated by the upper index of 
the $h$-functions, produced in 
the same initial partonic 
interaction; $\langle N_\hg \rangle$
denotes the mean number of hidden sources per collision. Note that 
 the $h$-functions include the angular dependence due to all possible 
correlations, namely, 
particle correlations in clusters, cluster correlations and hidden source 
correlations.

The bigger 
long-range correlations for the 3-particle correlation function
$C_3(\Delta \phi_{12},\Delta \phi_{13})$
in a 3-step cascade
are given by the $h^{(3)}(\Delta \phi_{12},\Delta
\phi_{13})$ term, 
 associated to three initial hidden/dark particles:
\beq\label{eq:h3}
h^{\rm (3)}(\Delta \phi_{12},\Delta \phi_{13})\ \sim\  
\exp{\left[-\frac{(\Delta \phi_{12})^2+(\Delta \phi_{13})^2-\Delta 
\phi_{12}\Delta \phi_{13}}
 {3\delta_{\hg \phi}^2+\delta_{\hg \phi}^2}\right]}
\eeq
\[
+\
\exp{\left[-\frac{(\Delta \phi_{12})^2}{2(2\delta_{\clg  
\phi}^2+\delta_{\hg \phi}^2)}\right]}+
\exp{\left[-\frac{(\Delta \phi_{13})^2}{2(2\delta_{\clg  
\phi}^2+\delta_{\hg \phi}^2)}\right]}+ 
\exp{\left[-\frac{(\Delta \phi_{12})^2+(\Delta \phi_{13})^2-2\Delta 
\phi_{12}\Delta \phi_{13}}{2(2\delta_{\clg \phi}^2+\delta_{\hg 
\phi}^2)}\right]}\, .
\]
Here, 
 $\delta_{\hg \phi}$ and $\delta_{\clg \phi}$ represent 
the expected 
correlation length due to the first
and second steps in the evolution of the parton cascade using a 
simplified model. In turn, correlations of particles from clusters
are parametrized by $\delta_{\phi}$, 
which can be referred to as the
cluster decay width in the transverse plane (see 
\cite{Sanchis-Lozano:2018wpz,Sanchis-Lozano:2016qdaw}).
The full set of expressions for the 3-particle correlation 
function
$C_3(\Delta \phi_{12},\Delta \phi_{13})$  
in a 3-step cascade process can be found in \cite{Sanchis-Lozano:2018wpz}.

 The plots of the 3-particle correlation function 
obtained from numerical estimates of a
3-step cascade with an initial long-range contribution from a hidden 
sector are shown in 
Fig.~\ref{fig:contour-proj-phi}.  The left panel shows a typical 
(spiderweb) structure of the 
3-particle 
correlations in a 2-dimensional ($\Delta \phi_{12}$, $\Delta 
\phi_{13}$)
plot.  The right panel shows its projection along the diagonal
$\Delta \phi_{12}=\Delta \phi_{13}$. The latter plot 
stands for 3-particle 
correlations disentangling the diverse sources of short-range 
and 
long-range
angular correlations.

\begin{figure}[t!]
\begin{center}
\includegraphics[scale=0.69]{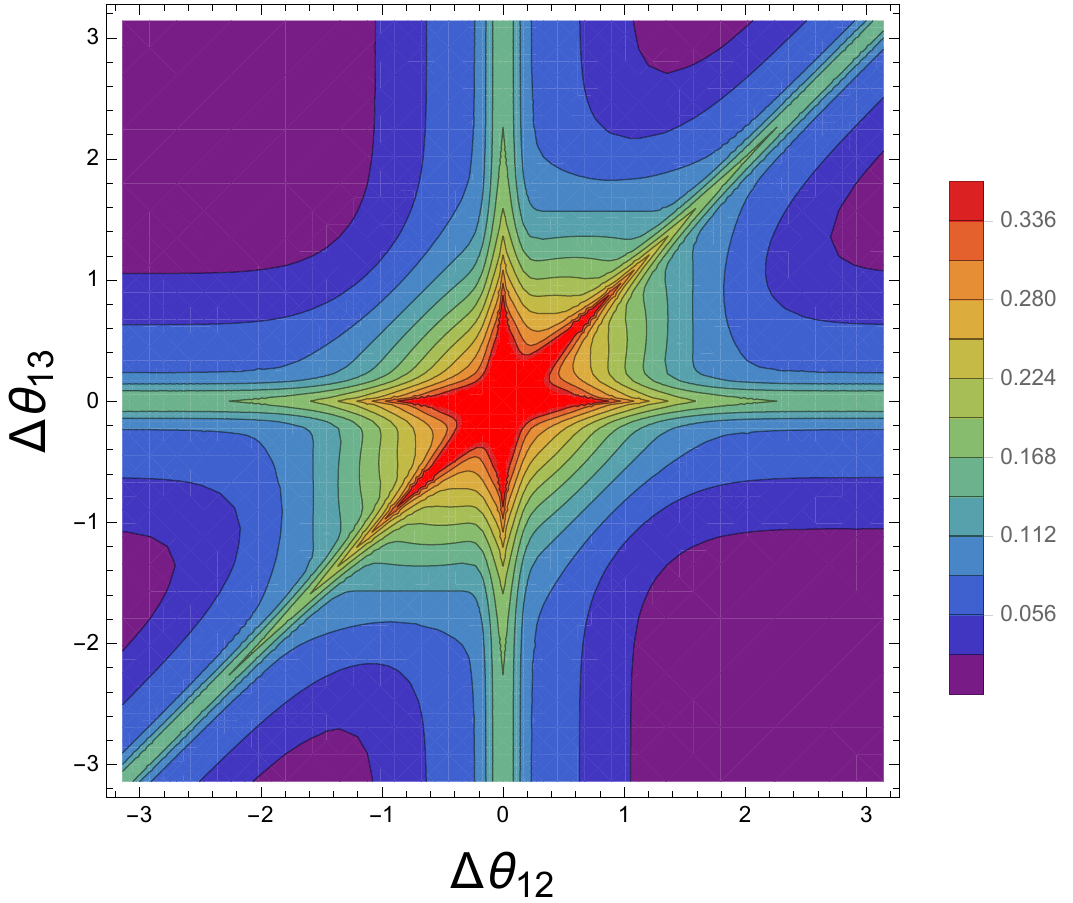}
\includegraphics[scale=0.69]{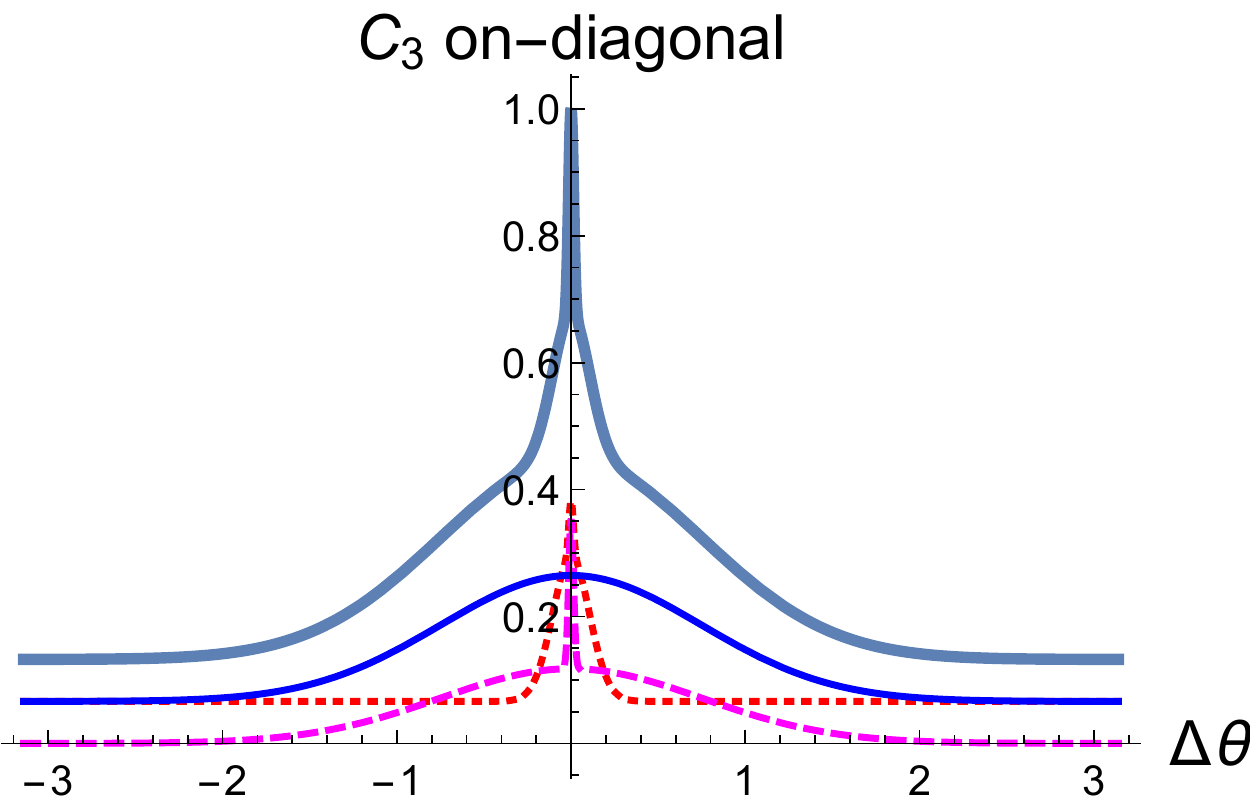}
\caption{{\em Left}: Contour plot 
of the 3-point correlation function  $C_3(\Delta 
\theta_{12},\Delta \theta_{13})$
in the 
cosmological case
using a toy model to take into account different sources of short-range 
and long-range correlations; 
{\em Right}: Diagonal projection of  $C_3(\Delta
\theta_{12},\Delta \theta_{13})$ 
showing the different contributions in analogy to Fig.2.} 
\label{fig:contour-proj-theta}
\end{center}
\vspace*{-0.3cm}
\end{figure}


Turning to the cosmological analogy, we apply the same analysis of angular 
correlations in the CMB as employed in high-energy collisions with the 
following correspondence in our Gaussian parametrization: $\delta_{\hg\phi} \to 
  \delta_{\hg\theta}$ for long-range correlations, $\delta_{\clg\phi} \to 
   \delta_{\clg\theta}$ for short-range 
correlations, and $\delta_{\phi} \to 
  \delta_{\theta}$
for very-short-range (within-cluster) correlations, 
respectively.

As commented above, 
  two  kinds of anisotropies can be 
  distinguished in 
the CMB:
   \begin{itemize}
 \item[(1)] 
	{\em Primary} anisotropies produced prior to recombination/decoupling, 
yielding rather long-range correlations mainly due to the SW effect. In 
our 
parametrization it corresponds to $\delta_{\hg\theta} \simeq 1$~rad (see 
Eq. 
(\ref{eq:thetamaxfinal})). Also, baryon acoustic oscillations of the 
plasma contribute 
to anisotropies but at a smaller scale, implying  $\delta_{\clg\theta} 
\lesssim 
1^\circ$.
 \item[(2)]
	{\em Secondary} anisotropies developing as the CMB propagates 
from the last 
scattering surface to the present observer leading to very-short-range 
angular 
correlations. The set of such effects yields in our parametrization to 
$\delta_{\theta} \ll 1^\circ$.
 \end{itemize}

Notice that the angular scales showing up 
as anisotropies 
in the CBM are not too much different from the expected azimuthal scales 
stemming from multiparticle production via
a hidden sector on top of the partonic shower in high-energy collisions
\cite{Sanchis-Lozano:2018wpz}.


Figure~\ref{fig:contour-proj-theta} shows the 3-point correlation function 
$C_3(\Delta \theta_{12},\Delta \theta_{13})$ plots for cosmological 
estimates using a simple model which takes into account the overall 
sources of short-range and long-range correlations. The 
left panel shows the 2-dimensional plot of $C_3(\Delta 
\theta_{12},\Delta \theta_{13})$
as a function of $\Delta \theta_{12}$ and 
 $\Delta \theta_{13}$ while the right panel shows the on-diagonal 
   projection of the function. This figure is analogous to Fig.2
 for high-energy collisions. 
 Again, 
 the 3-point correlation function is arbitrarily normalized to unity at 
$\Delta \theta_{12}=\Delta \theta_{13}=0$ 
since we are here interested rather in disentangling 
the different sources to angular correlations. 
Of course, 
a more realistic study should incorporate the absolute 
normalisation and relative weights using a more detailed model.  
Note that the $h(\Delta\theta_{12},\Delta\theta_{13})$ 
functions, i.e. the equivalent cosmological terms in Eq.(10) 
for high-energy collisions, are similarly
sensitive to very-short-range, short-range and long-range 
correlations, respectively. 
This is an important result of our work.

By comparing Figs.~\ref{fig:contour-proj-phi} and 
\ref{fig:contour-proj-theta}, 
an equivalent structure can be appreciated in both panels
as expected from the common existence of angular short-range 
and long-range
   correlations, no matter their physical origin. 
 Again, as in the case of high-energy hadronic collisions,  
  several 
distinct correlation scales clearly show up: short-range 
correlation lengths (secondary angular correlations),  
and a long-range correlation length (primary angular correlations)
associated to the early epoch of the universe. The 
   diagonal projection 
   suggests a pattern
which might be useful to disentangle possible sources of 
angular correlations present in the CMB. Short-range and very-short-range
correlations are behind 
the peak structure while 
longer correlations
determine the smooth falling off. Further
detailed structure can vary depending on the different underlying effects, 
  but the overall behaviour is expected to
be quite similar.

\section{Conclusions}

In this paper, we discuss  an intriguing similarity 
between long-range angular 
correlations observed in the CMB and 
 those obtained from multiparticle production in high-energy collisions. 
Although the physical origin of such long-range angular correlations is 
completely different in the two physical situations, the analogy is 
supported by the following facts: the time evolution in both cases 
(yielding complex structures from a primitive state of matter, either 
galaxies or final-state particles) is not continuous but rather involves 
different well-defined steps, with similar angular scales.
 Based on this observation 
 a common explanation has been proposed 
upon the assumption of the 
 existence of an unconventional early state: 
 an expanding universe 
 before recombination/decoupling (last scattering surface), 
where the CMB was 
released,
evolving up to present days, versus
 the formation of hidden/dark states in hadronic collisions followed by 
a conventional QCD cascade 
  resulting 
 in
  final-state particles.  
 Using 
  simple 
modeling, we show that 3-point/3-particle correlations should 
be a useful tool to disentangle the different contributions to 
short-range and 
long-range correlations in the universe evolution 
or in multiparticle production, highlighting deep 
connections between both fields in the search for new physics
and phenomena either  at the LHC or future accelerators.

\subsection*{Acknowledgments}
This work has been partially supported by the Spanish 
Ministerio de Ciencia, Innovaci\'on y Universidades, 
under grant FPA2017-84543-P and by Generalitat Valenciana
under grant PROMETEO/2019/113 (EXPEDITE). N.S-G 
is supported by the Funda\c c\~ao para a Ci\^encia e a Tecnologia (FCT) projects
PTDC/FIS-OUT/28407/2017 and UID/FIS/00099/2020 (CENTRA), CERN/FIS-PAR/0027/2019, and by 
the European  Union's  Horizon  2020  research  and  innovation (RISE) programme
H2020-MSCA-RISE-2017 Grant No.~FunFiCO-777740.
\vskip 0.1cm



\end{document}